\documentstyle[twocolumn,aps,epsf,floats]{revtex}

\begin{document}
\twocolumn[\hsize\textwidth\columnwidth\hsize\csname @twocolumnfalse\endcsname

\title{Luttinger theorem for a spin-density-wave state}
\author{B. L. Altshuler$^{1}$, 
A. V. Chubukov$^{2}$, A. Dashevskii$^{3}$, A. M. Finkel'stein$^3$ and D. K.
Morr$^2$}
\address{
$^{1}$ NEC Research Institute and Dept. of Physics, Princeton University, 
Princeton, NJ\\
$^{2}$Department of Physics, University of Wisconsin, Madison, WI 53706\\
$^{3}$ Weizmann Institute of Science, Rehovot 76100, Israel}
\date{\today}
\maketitle
\begin{abstract}
We obtained the analog of the Luttinger relation for a commensurate
spin-density-wave state. We show that while the relation between the area of
the occupied states and the density of particles gets modified in a simple
and predictable way when the system becomes ordered, a perturbative
consideration of the Luttinger theorem does not work due to the presence of
an anomaly similar to the chiral anomaly in quantum electrodynamics.
\end{abstract}
]

One of the key open problems in the study of high-$T_c$ superconductors is
the shape and the area of the electronic Fermi surface near half-filling.
Photoemission studies \cite{Gofron} performed near optimal doping have
demonstrated that the area enclosed by the Fermi surface is large, and for a
doping concentration $x$, constitutes a fraction $(1-x)/2$ of the area of
the Brillouin zone, just as for free fermions.  This is consistent with the
Luttinger theorem which states that in any conventional Fermi liquid, the
area enclosed by the Fermi surface does not change 
due to an interaction between fermions.  
On the other hand, similar experiments on underdoped materials have shown no
evidence of a Fermi surface crossing for momenta far from the zone diagonal~%
\cite{Shen}. These data are consistent with the idea that the Fermi surface
evolves with decreasing doping towards a small hole Fermi surface consisting
of four pockets centered at $(\pm \pi /2,\pm \pi /2)$. The existence of such
a Fermi surface in the paramagnetic phase would imply a violation of the
Luttinger theorem.

To consider a possible violation of the Luttinger theorem in a system with
strong magnetic fluctuations, one first needs to understand how this theorem
is modified when the system acquires a long-range magnetic order. This is
the issue which we address in this paper. We show that there are several
subtleties which emerge already on the mean-field level. In particular, a
naive perturbative approach yields incorrect results because of a hidden
anomaly which requires a proper regularization.

We begin with a brief consideration of the Luttinger theorem for a
conventional Fermi liquid~\cite{LuttWard}. The number of particles $N$ can
be expressed through the single particle Green function $G_k(\omega )$ of
the interacting electrons as  
\begin{equation}
N=-i~{\rm Tr}\sum_k~\lim_{t\rightarrow +0}\int \frac{d\omega }{(2\pi )}%
G_k(\omega )e^{i\omega t},  \label{defG}
\end{equation}
where ${\rm \ Tr}$ implies a summation over spin projections. Using the %
obvious relation for $G$ and the self-energy $\Sigma $, $G_k(\omega
)=~\partial \log {G_k^{-1}(\omega )}/\partial \omega +G_k(\omega )\partial
\Sigma _k(\omega )/\partial \omega $, one can rewrite Eq.(\ref{defG}) as $%
N=I_1-I_2$ where 
\begin{eqnarray}
I_1 &=&-i~{\rm Tr}\sum_k~\int_{-\infty }^\infty \frac{d\omega }{(2\pi )}~%
\frac \partial {\partial \omega }\log G_k^{-1}(\omega ),  \nonumber \\
I_2 &=&i~{\rm Tr}\sum_k~\int_{-\infty }^\infty \frac{d\omega }{(2\pi )}%
G_k(\omega )~\frac \partial {\partial \omega }\Sigma _k(\omega ).  \label{I}
\end{eqnarray}
In Ref.\cite{LuttWard}, Luttinger and Ward (LW) argued that $I_2=0$ to all
orders in perturbation theory around free fermions. %
Their consideration is based on a functional $Y=-i~{\rm Tr}~\sum_k~\int
d\omega /(2\pi )~Y_k(\omega )$ determined by the variational equation 
\begin{equation}
\delta Y=-i~{\rm Tr}~\sum_k~\int\frac {d\omega}{2\pi}~\Sigma_k(\omega)~\delta
G_k(\omega). 
\label{vareq}
\end{equation}
This variational equation 
allows one to re-express $I_2$ as an integral of a full derivative: 
\begin{equation}
I_2=-i~{\rm Tr}\sum_k~\int_{-\infty }^\infty \frac{d\omega }{(2\pi )}\frac{%
\partial Y_k(\omega )}{\partial \omega }.  \label{bbb}
\end{equation}
LW gave a recipe how to obtain the functional $Y_k(\omega )$ order by order
in a diagrammatic perturbation theory. The perturbative $Y_k(\omega )$ is
free from singularities and 
vanishes at large frequencies.
Obviously, in this situation $I_2=0$.
(We however will demonstrate below that 
$I_2$ may be finite due to nonperturbative effects). 
The term $I_1$ also contains a full derivative over frequency and,
apparently, should vanish for
similar reasons. However, its calculation requires care as $\log G$ is a
non-analytic function of frequency in the upper half-plane. A careful
consideration shows that due to a nonanalyticity, $I_1$ turns out to be
finite: 
\begin{equation}
I_1=\frac 1{2\pi i}~{\rm Tr}\sum_k\log \frac{G_k^{-1}(\mu +i0)}{G_k^{-1}(\mu
-i0)},  \label{I1}
\end{equation}
where $\mu $ is the chemical potential of the interacting electrons. 
It then follows that 
\begin{equation}
N=I_1={\rm Tr}~\sum_k\Theta 
(E_k-\mu),  \label{LTconv}
\end{equation}
where $E_k$ is the quasiparticle energy, and the $%
\Theta $-function counts the states inside the Fermi surface. This is the
Luttinger theorem for a conventional Fermi liquid.  
It states that the number of states enclosed by the Fermi surface is equal
to the number of particles $N$. 

The Luttinger relation can be easily extended to the case when the system
develops a long-range magnetic order. Consider for definiteness a 2D
spin-fermion model which describes a system of propagating fermions with
dispersion $\epsilon _k$ coupled to localized spins $S_q$ by an exchange
interaction \cite{MP,ChubMorr}: ${\cal H}_{int}=g\sum_{{\bf q,k},\alpha
,\beta }c_{\alpha ,k+q}^{\dagger }{\vec{\sigma}}_{\alpha \beta }c_{\beta ,k}~%
{\vec{S}_q}$. Here $\sigma ^i$ are the Pauli matrices, and the localized
spins are described by their dynamical spin susceptibility, $\chi (q,\omega )
$. For simplicity, we assume that there are no other interactions, i.e., for 
$g=0$ one deals with free fermions. In the antiferromagnetically ordered
phase one of the components of ${\vec{S}}$ acquires a nonzero expectation
value $<S_q^z>=<S^z>\delta _{q,Q}$. In the mean-field spin-density-wave
(SDW) theory, 
the interaction term reduces to $\Delta c_{\alpha ,k+Q}^{\dagger }\sigma
_{\alpha \alpha }^zc_{\alpha ,k}$, where $\Delta =g<S^z>$. Problems of this
kind are most conveniently described in a matrix formalism. In the presence
of the SDW order, the electronic spectrum can be obtained from the $4\times 4
$ matrix 
\begin{equation}
G_k^{-1}(\omega )=\left[ 
\begin{array}{cc}
\omega -{\bar{\epsilon}}_k & \Delta \,\,\sigma _z \\ 
\Delta \,\,\sigma _z & \omega -{\bar{\epsilon}}_{k+Q}
\end{array}
\right] \text{,}  \label{matrix}
\end{equation}
where ${\bar{\epsilon}}_k=
\epsilon _k-\mu$, and $\mu$ is the
actual, $\Delta$-dependent chemical potential.
 For the purpose of comparison
with the paramagnetic phase we will 
keep using the original Brillouin zone 
rather than 
the magnetic one. 
Then, for each $k$ one gets two
solutions ${\bar{E}}_k^{~\pm ~}$ $=({\bar{\epsilon}}_k+{\bar{\epsilon}}%
_{k+Q})/2~\pm ~[\Delta ^2+(({\bar{\epsilon}}_k-{\bar{\epsilon}}%
_{k+Q})/2)^2]^{1/2}$ which satisfy ${\bar{E}}_k^{\pm }={\bar{E}}_{k+Q}^{\pm }
$ . The relations (\ref{defG}),(\ref{I}) between $N$ and the single particle
Green's function 
still hold in the ordered state, with the only modification that ${\rm Tr}$
now acts on 
$4 \times 4 $
matrices. In a mean-field approximation the $I_2$ term vanishes identically,
and hence $N=I_1$. 
A substitution of Eq.(\ref{matrix}) into Eq.(\ref{I1})  yields
$N=\sum_k\left( \Theta ({\bar{E}}_k^{+})+\Theta ({\bar{E}}%
_k^{-})\right).$
Subtracting this relation from the total number of states in the Brillouin
zone, $N_{BZ}$, one finally obtains 
\begin{equation}
x=S_{hole}-S_{double},  \label{area}
\end{equation}
where $x\equiv 1-N/N_{BZ}$ is a hole doping concentration, while $S_{hole}$
and $S_{double}$ are fractions of the Brillouin zone occupied by hole
pockets (${\bar{E}}^{\pm }>0$) and doubly occupied electronic states (${\bar{%
E}}^{\pm }<0$), respectively. Eq.(\ref{area}) is the Luttinger relation for
the ordered SDW phase. It has been derived here in a mean-field
approximation, but Eq.(\ref{area}) should also hold in the presence of
fluctuations due to an
interaction between fermions, simply because the 
effect of the spin ordering is completely absorbed into the matrix
formulation of the LW functional $Y$. 

\begin{figure}[t]
\begin{center}
\leavevmode
\epsffile{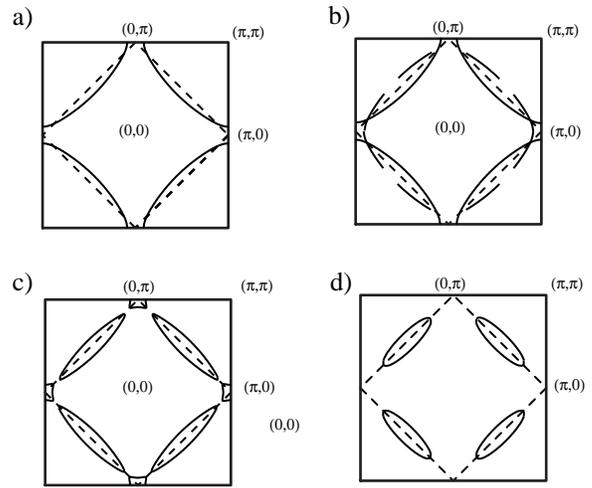}
\end{center}
\caption{Schematic evolution of the Fermi surface with the gap $\Delta $ in
the ordered SDW phase with electron hopping to nearest ($t$) and
next-nearest($t^{\prime })$ neighbors on a square lattice. 
We set $
t^{\prime }=-0.2t$. The dashed line indicates the boundary of the magnetic
Brillouin zone.}
\label{fsord}
\end{figure}

We now give the physical
interpretation of the Luttinger relation for the ordered state. For this, we
fix the doping concentration at some small but finite level and follow the
Fermi surface evolution with increasing $\Delta $. For $\Delta =0$, the
Fermi surface has the form shown in Fig~\ref{fsord}a - it is centered at $%
(\pi ,\pi )$ and encloses an area which is slightly larger than half of the
Brillouin zone, in accordance with the conventional Luttinger theorem, Eq. (%
\ref{LTconv}). 
Switching on an infinitesimally small $\Delta $ doubles the unit cell in
real space. As a result, extra pieces of the Fermi surface appear because a
Fermi surface crossing at $k$ necessarily implies one at $k+Q$ (Fig.~\ref
{fsord}b). The subsequent evolution of the Fermi surface proceeds as is
shown in Fig.~\ref{fsord}c,d. For sufficiently large $\Delta $, the doubly
occupied electron states disappear, and the Fermi surface consists of just
four small hole pockets centered at $(\pi /2,\pi /2)$ (see Fig.~\ref{fsord}%
d). The total area enclosed by the pockets is small and scales as $x$ in
agreement with Eq.(\ref{area}).

So far, the LW argumentation works perfectly. However, the matrix formalism
cannot be extended to the paramagnetic phase. We now discuss another,
complimentary approach to the ordered state in which one does not explicitly
introduce a condensate of $S^z$ but instead obtains the SDW form of the
electronic spectrum by introducing an exchange of $\delta -$functional
longitudinal spin fluctuations: $\chi _l(q,\Omega )=(\Delta /g)^2\delta
(\Omega )\delta (q-Q)$. This approach is advantageous as it can be applied
to the paramagnetic phase - one just has to substitute the $\delta -$%
functional longitudinal susceptibility by the one with a finite spin-spin
correlation length~ \cite{ChubMorr,KS}. 
In the spin-fluctuation formalism,
the full Green's function does not have a matrix structure, but there
appears a self-energy due to spin-fluctuation exchange. To the lowest order
in $\Delta ^2$, 
\begin{equation}
\Sigma_k(\omega )=\Delta ^2G_{k+Q}^{(0)}(\omega )=\frac{\Delta ^2}{\omega -{%
\bar{\epsilon}}_{k+Q}},  \label{sigma}
\end{equation}
where $G_k^{(0)}(\omega )=(\omega -{\bar{\epsilon}}_k)^{-1}$ is the Green's
function of free electrons. For the 
$\delta -$functional form of $\chi _l$ this is in fact an exact result - the
higher order terms in $\Sigma $ are absent due to a Ward identity \cite
{ChubMorr}. The full Green's function then takes the form 
\begin{equation}
G_k(\omega )=\frac{\omega -{\bar{\epsilon}}_{k+Q}}{(\omega -{\bar{E}}%
_k^{+})(\omega -{\bar{E}}_k^{-})}.  \label{G}
\end{equation}
Not surprisingly, it coincides with the diagonal part of the matrix $G$
defined by Eq.(\ref{matrix}). 
At the same time, the $I_{1,2}$ terms are not
 the same in the matrix and the diagonal
formalisms. Moreover, 
the self-energy in 
(\ref
{sigma}) has a singular frequency dependence.
We therefore have to reexamine
the argument that $I_2=0$. 
To begin with, we first observe that $I_1$ 
alone does
not yield the right expression for $N$.
 Indeed, it follows from Eqs.(\ref{I},\ref{G})
that 
\begin{equation}
I_1=~2\sum_k\left( \Theta ({\bar{E}}_k^{+})+\Theta ({\bar{E}}%
_k^{-})-\Theta ({\bar{\epsilon}}_{k+Q})\right).
\label{bbbb}
\end{equation}
For large $\Delta $, $%
I_1=2(1-x)N_{BZ}$ in clear disagreement with Eq.(\ref{area}). This implies
that $I_2$ should be finite.
Observe that the numerator in (\ref{G}) also contributes to $I_1$.

To calculate $I_2$, we use Eq.(\ref{bbb}) which relates $I_2$ to the
LW functional $Y$. This functional can be straightforwardly 
obtained by performing
 the functional integration of the self-energy $\Sigma$ over $G_k(\omega )$. To 
do this, we first have to 
re-express $\Sigma $ in terms of the full
Green's functions.
This can be done either directly, by substituting 
$G_{k+Q}^{(0)}(\omega )$ in
Eq.(\ref{sigma}) in terms of $G$, 
or diagrammatically, 
by rewriting
the self-energy as 
$\Sigma_k(\omega )=\Delta ^2G_{k+Q}(\omega )\Gamma_{k,k+Q}(\omega )$, 
and collecting diagrams for the full vertex $\Gamma $
order by order in the perturbative expansion, each time using full
propagators $G$ for the internal fermionic lines. It turns out, however,
that the results of the two approaches are  different, because there
exists a range of frequencies where the perturbative expansion is not
convergent. Indeed, a formal summation of the perturbation series for the
self-energy yields $\Sigma_k^{pert}(\omega )=2\Delta ^2G_{k+Q}(\omega
)/\left( 1+\sqrt{1+4\Delta ^2G_k(\omega )G_{k+Q}(\omega )}\right) $. On the
other hand, the use of an exact relation between
 bare and full Green's functions yields
\begin{equation}
\Sigma_k(\omega )=\frac{2\Delta ^2G_{k+Q}(\omega )}{1\pm \sqrt{1+4\Delta
^2G_k(\omega )G_{k+Q}(\omega )}}. 
 \label{SigmaG}
\end{equation}
Here the upper and lower signs should be used when $|y|<1$ and $|y|>1$,
respectively, where $y=\Delta ^2G_k^{(0)}(\omega )G_{k+Q}^{(0)}(\omega )$.
 Clearly,
the perturbative and exact expressions for the self-energy coincide only for $%
|y|<1$, i.e., when the perturbative expansion is convergent. 

We now show
that due to the difference between $\Sigma $ and $\Sigma ^{pert}$, the
actual $I_2$ is finite, while the perturbative approach yields $I_2=0$ to
all orders in $\Delta $. 
Indeed,
performing the integration over the Green's function in Eq.(\ref{vareq})
and using Eq.(\ref{SigmaG}) for $\Sigma$, we find 
\begin{eqnarray}
Y_k(\omega ) &=&\frac 12\Bigg[-\log G_k(\omega )\pm \sqrt{1+4\Delta
^2G_k(\omega )G_{k+Q}(\omega )}  \nonumber \\
&\pm &\frac 12\log \Bigg( \frac{\sqrt{1+4\Delta ^2G_k(\omega )G_{k+Q}(\omega )}-1}{%
\sqrt{1+4\Delta ^2G_k(\omega )G_{k+Q}(\omega )}+1}\Bigg) \Bigg].  \label{YY}
\end{eqnarray}
After simple manipulations, $Y_k(\omega )$ can be rewritten as
\begin{equation}
Y_k(\omega )=\frac 14\log \frac{G_k^{(0)}(\omega
)G_{k+Q}^{(0)}(\omega)}{G_k^2(\omega)}+Y_k^{reg}(\omega ),
\label{YYY}
\end{equation}
where $Y_k^{reg}(\omega )$ is a regular function of frequency.
We now substitute the functional $Y_k(\omega)$ into Eq.(\ref{bbb})
for $I_2$.
The integral 
in 
(\ref{bbb}) contains a full derivative but, similarly to what we 
previously obtained
in the calculation of $I_1$, it does not vanish. 
The point is that 
due to a nonanaliticity, 
the 
logarithmic term in Eq.(\ref{YYY}) 
does not allow one to reduce the integral to just the
values of $Y_k(\omega)$ at the limits of 
integration. 
Doing the same manipulations as with $I_1$, we obtain
\begin{equation}
I_2=\sum_k\left( \Theta ({\bar{E}}_k^{+})+\Theta ({\bar{E}}_k^{-})-2\Theta ({%
\bar{\epsilon}}_k)\right) \;.  \label{I_2}
\end{equation}
Collecting now the $I_1$ and $I_2$ contributions to $N$
we indeed reproduce
the correct result, Eq.(\ref{area}). It is essential, however, that $I_2$ is
finite (e.g., for large $\Delta $, $I_2=(1-x)N_{BZ}$ ). On the other hand,
we have checked that if we use the perturbative 
expression
for $\Sigma $, we
obtain $I_2$ in the form
of an integral of a full derivative over frequency of a
regular function $Y_k^{pert}(\omega )$ which vanishes at $\omega =\pm \infty 
$. In this case, the integral is indeed equal to zero.

To make this point clearer 
 and also to discuss some analogy with field
theory, we now compute $I_2$ directly by substituting the expressions for
the full Green's function and the full self-energy into Eq.(\ref{I}). We
find 
\begin{eqnarray}
I_2&=&-2i \Delta^2 \sum_k \int \frac{d\omega}{2\pi}~
(\omega -\bar{{\epsilon}}%
_{k+Q}+i\delta _\omega )^{-1}
\nonumber \\
&&\times [(\omega -\bar{E}_k^++i\delta _\omega
)(\omega -\bar{{E}}_k^-+i\delta _\omega )]^{-1},
\label{nnn}
\end{eqnarray}
 where $\delta _\omega =\delta \ sgn \omega$.
In a perturbation theory, one assumes that $\Delta $ is small and expands $I_2$
in powers of $\Delta $: $I_2=\sum_{n=1}^\infty \Delta ^{2n}I_2^{(n)}$, where $%
I_2^{(n)}$ is evaluated at $\Delta \equiv 0$. 
Performing  
this expansion in (\ref{nnn}) and integrating over frequency,
we find 
\begin{equation}
I_2^{(n)}= \sum_k~~\frac{\Theta ({\bar{\epsilon}} _k)-\Theta ({\bar{\epsilon}%
}_{k+Q})}{({\bar{\epsilon}}_k-{\bar{\epsilon}} _{k+Q})^{2n}}.
\label{perturb}
\end{equation}
Clearly, $I_2^{(n)}=0$ for all $n$ by symmetry: the summation over $k$ in (%
\ref{perturb}) goes over the full Brillouin zone while the function changes
sign when $k$ and $k+Q$ are interchanged. In principle, the very fact that $%
I_2^{(n)}=0$ to all orders in perturbation theory does not exclude an
exponential dependence of $I_2$ on $1/\Delta ^2$. However, we explicitly
computed $I_2$ numerically using Eq.(\ref{I_2}) and found that $I_2\propto
\Delta ^2$ at small $\Delta$, i.e., $I_2$ possesses a regular dependence on
the expansion parameter. For $t^{\prime }=-0.2t$ and $\mu =-0.69t$, we found 
$I_2\approx -0.63\Delta ^2/t^2$.

We now show that the reason for the discrepancy between the explicit and
perturbative calculations lies in the fact that $lim_{\Delta \rightarrow
0}\;I_2/\Delta ^2\neq I_2^{(1)}$ where $I_2^{(1)}$ is given by Eq.(\ref
{perturb}). Indeed, examine the form of $I_2^{(1)}$ before the frequency
integration is performed. We have 
$$I_2^{(1)}=-2i\sum_k \int \frac{d\omega}{2\pi}
\frac{1}{(\omega -{\bar{\epsilon}}_{k+Q}+i\delta _\omega )^2~(\omega -{\bar{
\epsilon}}_k+i\delta _\omega )}.
$$
 We see that the integrand in $%
I_2^{(1)}$ contains a double pole at $\omega =\bar{\epsilon}_{k+Q}$, i.e.,
the frequency integral is in fact linearly divergent. This divergence,
however, does not show up in the final result as changing $k$ to $k+Q$, one
finds the same divergence, but with opposite sign. Compare now this formula
with the exact (nonperturbative) form for $I_2$, Eq.(\ref{nnn}).
We see that the inclusion of the SDW self-energy correction into
the Green's function splits the double pole at $\omega ={\bar{\epsilon}}%
_{k+Q}$ into two closely located poles at $\omega ={\ \bar{\epsilon}}_{k+Q}$
and $\omega =\bar{E}_k^-$ for $\epsilon _k>\epsilon _{k+Q}$, or $\omega =\bar{%
E}_k^+$ for $\epsilon _k<\epsilon _{k+Q}$. For most of the Brillouin zone,
both poles lie in the same half-plane. However, for $\bar{ \epsilon}_k\bar{%
\epsilon}_{k+Q}<\Delta ^2$, the poles at ${\bar{\epsilon}} _{k+Q}$ and,
e.g., $\bar{E}_k^+$ are located in different half-planes in which case the
frequency integration yields a contribution which is absent in the case $%
\Delta=0$. The region in momentum space where this new contribution exists
is very narrow, its area is of the order of $\Delta ^2$. However, everywhere
in this region, the denominator in $I_2$ is also of the order of $\Delta ^2$
as $\bar{E}_k^+-{\bar{\epsilon}}_{k+Q}\sim \Delta ^2$. As a result, the
additional contribution to $I_2/\Delta ^2$ remains finite as $\Delta
\rightarrow 0$. 
To calculate this contribution,
 we transformed the summation over momenta into the
integration over energies and integrated over a narrow region in ${\ \bar{%
\epsilon}}_{k+Q}:
|{\bar{\epsilon}}_{k+Q}|<\Delta ^2/|{\bar{\epsilon}} _k|$.
We then obtained $I_2/\Delta ^2$ in the form of a one-dimensional integral over $%
d{\bar{\epsilon}}_k$: 
\begin{equation}
lim_{\Delta \rightarrow 0} \; I_2/\Delta^2 = \frac{1}{4 \pi^2 t
|t^{\prime}|} \int^{{\bar\epsilon}^{max}}_{{\bar\epsilon}^{min}} \frac{d{\bar%
\epsilon}_k}{{\bar\epsilon}_k W({\bar\epsilon}_k)},  \label{result}
\end{equation}
where $W({\bar\epsilon}_k) = ((1+a)^2 -b^2)(b^2 -4a))^{1/2},~a = ({\bar%
\epsilon}_k -2|\mu|)/(8|t^{\prime}|),~ b = {\bar\epsilon}_k/(4t), ~{\bar%
\epsilon}^{max} = 4|\mu|/(1 + (1-|\mu t^{\prime}|/t^2)^{1/2}),~ {\bar\epsilon%
}^{min} = 2(4t^{\prime} - \mu)/(1 + 2 |t^{\prime}|/t) <0$. Performing the
one-dimensional integration numerically, we found for $t^{\prime} =-0.2t$
and $\mu =-0.69 t$, $I_2 \approx - 0.63 \Delta^2/t^2$ which is 
exactly the same result that we obtained using Eq.(\ref{I_2}). We have
checked that this equivalence also holds for other values of $t^{\prime}$
and $\mu$. Notice that contrary to naive expectations, the integration in
Eq.(\ref{result}) is not confined to a region near the points where the
Fermi surface crosses the magnetic Brillouin zone boundary (hot spots).
These points correspond to ${\bar\epsilon}_k=0$ in the integrand in Eq.(\ref
{result}). In fact, if we substitute $W({\bar\epsilon}_k)$ by $W(0)$, we
would obtain a very different $I_2$ which for some $t^{\prime}$ even possesses
the wrong sign.

The anomaly which gives rise to a finite value of $I_2$ is very similar to
the chiral anomaly in quantum electrodynamics \cite{Jackiw}. In both cases,
we have an integral which apparently vanishes by symmetry, but contains a
hidden linear divergence. Performing a regularization, one gets rid of the
divergence but simultaneously brakes the symmetry. In our case, the proper
regularization is achieved by performing all computations at a small but
finite $\Delta$, i.e., keeping the symmetry broken at all stages of the
calculations. After regularization, the double pole gets  
split into two
poles, and there appears an extra contribution from the region, where the
poles are in different half-planes. This extra contribution turns out to be
finite because the smallness of the phase space is fully compensated by the
smallness of the energy denominator.

Finally, let us discuss how the opening of the gap $\Delta$ influences the 
chemical potential $\mu$. Combining Eqs.(\ref{area}) and (\ref{bbb}), we 
find $I_2=2 \sum_k[\Theta(\epsilon_k-\mu_{\Delta =
0})-\Theta(\epsilon_k-\mu)]$. Clearly, the fact that $I_2$ is finite implies 
that $\mu \neq \mu_{\Delta=0}$. For large $\Delta$, $|\mu|\approx \Delta \gg 
|\mu_{\Delta=0}|$.

In summary, in this paper we obtained the Luttinger relation for the
antiferromagnetically ordered state. For a formalism where one does not
introduce an antiferromagnetic order parameter, we have demonstrated that
the perturbative derivation of the Luttinger theorem breaks down because of
a hidden anomaly which needs to be regularized. We found similarities
between this phenomenon and the chiral anomaly in quantum electrodynamics.

The present paper only considers the magnetically ordered state. It was
recently argued by two of us \cite{ChubMorr} that $I_2$ remains finite even
in the paramagnetic phase if the spin-fermion coupling strength exceeds some
threshold value. This argument is consistent with the idea of small hole
pockets in heavily underdoped cuprates. A more detailed study of the
possible violation of the Luttinger theorem in the paramagnetic phase is
clearly called for.

It is our pleasure to thank all colleagues with whom we discussed this
issue. This research was supported by NSF DMR-9629839 (for A. C. and D. M.),
and by the German-Israel Foundation - GIF (for A. D. and A. F.). A. Ch. is
an A.P. Sloan fellow. A. F. is supported by the Barecha Fund Award.

\end{document}